\documentclass[12pt]{article}
\usepackage[margin=1in]{geometry}
\usepackage{amsmath,amssymb}
\usepackage{graphicx}
\usepackage{url}

\title{BoolForge: Controlled Generation and Analysis of Boolean Functions and Networks}
\author{
    Claus Kadelka$^{1,*}$, Benjamin Coberly$^{1,2}$ \\
    \\
    $^1$Department of Mathematics, Iowa State University, Ames, IA, USA \\
    $^2$Department of Computer Science, Iowa State University, Ames, IA, USA \\
    *To whom correspondence should be addressed. Email: ckadelka@iastate.edu
}

\date{}

\begin{document}
\maketitle

\begin{abstract}

\textbf{Summary:}
Boolean networks are a widely used modeling framework in systems biology for studying gene regulation, signal transduction, and cellular decision-making. Empirical studies indicate that biological Boolean networks exhibit a high degree of canalization, a structural property of Boolean update rules that stabilizes dynamics and constrains state transitions. Despite its central role, existing software packages provide limited support for the systematic generation of Boolean functions and networks with prescribed canalization properties.

We present \texttt{BoolForge}, a Python toolbox for the random generation and analysis of Boolean functions and networks, with a particular focus on canalization. \texttt{BoolForge} enables users to (i) generate random Boolean functions with specified canalizing depth, layer structure, and related constraints; (ii) construct Boolean networks with tunable topological and functional properties; and (iii) analyze structural and dynamical features including canalization measures, robustness, modularity, and attractor structure. By enabling controlled generation alongside analysis, \texttt{BoolForge} facilitates ensemble-based investigations of structure-dynamics relationships, benchmarking of theoretical predictions, and construction of biologically informed null models for Boolean network studies.

\textbf{Availability and Implementation:}
BoolForge is implemented in Python ($\geq$3.10) and can be installed via \texttt{pip install boolforge}.
Source code and documentation are available at \url{https://github.com/ckadelka/BoolForge}. A PDF tutorial compendium is provided as Supplementary Material.
\end{abstract}

\section{Introduction}
Boolean network (BN) models are widely used to study qualitative aspects of regulation, decision-making, and information processing in biological systems~\cite{wang2012boolean,hemedan2022boolean}. From Kauffman’s early work on genetic regulatory circuits~\cite{KAUFFMANfirstpaper} to recent large-scale and data-driven applications~\cite{zerrouk2024large}, Boolean networks provide an interpretable and computationally tractable representation of complex dynamical systems.

Empirical studies of biological Boolean network models consistently reveal strong \emph{canalization} in their update rules -- a hierarchical and redundancy-rich organization of input variables~\cite{daniels2018criticality,gates2021effective,kadelka2024meta}. Canalization is thought to contribute to the stability, robustness, and evolvability of living systems. Theoretical advances have characterized canalizing depth, layer structure, collective canalization, and input redundancy~\cite{he2016stratification,kadelka2017influence,dimitrova2022revealing}, providing a detailed mathematical understanding of this functional architecture. However, software support for systematically generating Boolean functions or networks with prescribed canalization properties remains limited.

Most existing tools for Boolean network analysis, including \texttt{BoolNet}~\cite{mussel2010boolnet}, \texttt{PyBoolNet}~\cite{klarner2017pyboolnet}, \texttt{Cyclone}~\cite{dimitrova2023cyclone}, and \texttt{biobalm}~\cite{trinh2025mapping}, primarily focus on simulation, attractor identification, and dynamical analysis. While highly effective for studying specific models, they provide little support for the systematic generation of Boolean functions or networks with prescribed canalization properties (e.g., sampling functions with fixed canalizing depth or layer structure). Yet the ability to generate controlled ensembles of Boolean functions and networks is essential for hypothesis testing, benchmarking, and investigating how structural features such as canalization, bias, and connectivity shape network behavior. Such ensembles provide biologically informed null models against which structural and dynamical properties of curated regulatory networks can be evaluated.

To address this gap, we introduce \texttt{BoolForge}, a Python toolbox for the controlled generation and analysis of Boolean functions and networks, with an emphasis on canalization and ensemble-based investigations. \texttt{BoolForge} provides (i) flexible sampling algorithms for generating Boolean functions with user-defined structural and functional properties, including canalizing depth, layer structure, bias, and redundancy; (ii) methods for constructing Boolean networks with tunable wiring diagrams and rule distributions; and (iii) tools for analyzing structural, modular, and dynamical features. By integrating controlled generation with analysis in a single framework, \texttt{BoolForge} enables systematic ensemble studies, biologically informed null-model construction, and rapid prototyping of Boolean network models. In the following sections, we describe the architecture and core functionality of \texttt{BoolForge} and illustrate its capabilities through representative ensemble-based applications.

\section{Architecture and core functionality}

\begin{figure}
    \centering
    \includegraphics[width=\linewidth]{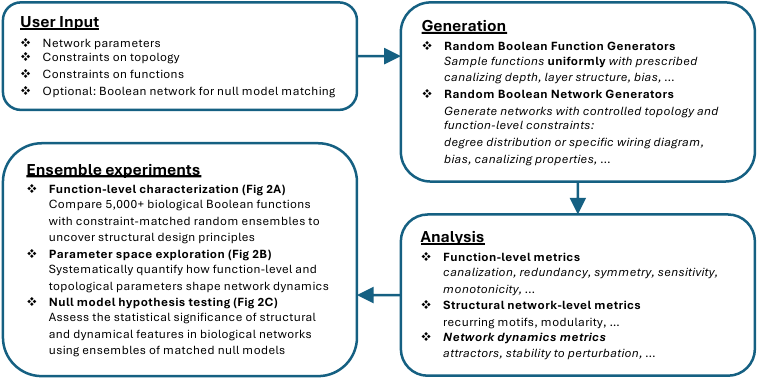}
    \caption{{\bf Overview of the \texttt{BoolForge} framework}. User-defined constraints on network topology and Boolean update rules guide the controlled generation of Boolean functions and networks. Integrated analysis tools quantify structural and dynamical properties and enable ensemble experiments such as parameter exploration and biologically informed null-model testing (examples in Fig.~\ref{fig:examples}).}
    \label{fig:workflow}
\end{figure}

\texttt{BoolForge} is designed to support systematic, ensemble-based investigation of Boolean functions and networks. Its architecture separates rule-level structure from network topology and integrates controlled random generation with structural and dynamical analysis. The framework centers on two primary abstractions, \texttt{BooleanFunction} and \texttt{BooleanNetwork}, enabling independent manipulation of update rules and wiring diagrams while maintaining interoperability with existing software (Fig.~\ref{fig:workflow}).

\subsection{Design and architecture}

\texttt{BoolForge} is organized around three guiding design principles. 
First, rule structure and network topology are treated as conceptually distinct objects. 
A \texttt{BooleanFunction} represents a local update rule independently of its network context, while a \texttt{BooleanNetwork} combines such functions with an explicit wiring diagram. 
For example, a Boolean function can be instantiated directly from a logical expression:

\begin{verbatim}
from boolforge import BooleanFunction
f = BooleanFunction("x1 & !x2 | x3")
\end{verbatim}

Boolean networks are then constructed by combining such functions with a wiring diagram. 
This separation enables controlled experiments in which functional and topological constraints can be varied independently.

Second, random generation and analysis are integrated within a single framework. 
Sampling algorithms are implemented alongside structural and dynamical metrics, allowing users to generate constrained ensembles and immediately evaluate their properties without switching software environments.

Third, the package emphasizes extensibility and interoperability. 
Models can be constructed from truth tables or logical expressions, converted from external formats, and exported for downstream analysis in complementary tools.

\subsection{Controlled generation and analysis of Boolean functions}

While generators for some basic function classes are available in other libraries, \texttt{BoolForge} provides uniform generators for several structurally constrained function families, including non-degenerate Boolean functions, linear and affine functions~\cite{chandrasekhar2023stability}, functions with specified canalizing depth~\cite{he2016stratification}, nested canalizing functions~\cite{li2013boolean}, functions with prescribed canalizing layer structure~\cite{dimitrova2022revealing}, and functions with fixed Hamming weight or bias. These generators enable controlled sampling within precisely defined structural classes, facilitating hypothesis testing about the role of canalization, bias, or redundancy in network behavior.

In addition to generation, \texttt{BoolForge} provides comprehensive structural analysis of Boolean functions. It identifies essential and non-essential variables, monotonicity properties, and input symmetries, and computes measures such as average sensitivity. Canalization is quantified through multiple complementary frameworks, including extended monomial representations revealing canalizing layer structure~\cite{he2016stratification,dimitrova2022revealing}, canalizing strength~\cite{kadelka2023collectively}, and input redundancy and effective degree~\cite{gates2021effective,marcus2025cana}. Together, these tools provide a detailed characterization of the functional building blocks underlying Boolean networks.

\subsection{Controlled generation of Boolean networks and null models}

A \texttt{BooleanNetwork} consists of a collection of \texttt{BooleanFunction} objects together with a wiring diagram specifying regulatory interactions. Networks can be constructed manually, imported from standard formats (including \texttt{PyBoolNet} and \texttt{CANA}), or generated randomly.

Random network generation proceeds in two steps. First, a wiring diagram is constructed. Users may fix in-degrees, sample from specified in-degree distributions, enforce strong connectivity, allow or disallow self-loops, or provide custom wiring diagrams (e.g., from existing biological Boolean network models). Second, update rules are assigned to nodes using any of the constrained function generators described above. This two-step construction separates topological and functional constraints, allowing their independent manipulation and systematic study.

\texttt{BoolForge} also provides routines for generating biologically informed null models to support statistical hypothesis testing. These null ensembles preserve selected features of a reference network -- such as degree distribution, canalizing depth, layer structure, or bias -- while randomizing other aspects. Such matched ensembles enable rigorous comparison of observed structural or dynamical properties against expectations under controlled constraints.

\subsection{Structural and dynamical analysis of Boolean networks}

Once networks are constructed, \texttt{BoolForge} provides integrated tools for structural and dynamical analysis. Structural routines identify canonical motifs, including feedback and feed-forward loops~\cite{alon2007network,kadelka2024meta}, and support quantitative assessment of modular organization~\cite{kadelka2023modularity}.

For dynamical analysis, \texttt{BoolForge} supports attractor identification under both synchronous and asynchronous update schemes, robustness metrics, and measures of network, attractor, and basin coherence~\cite{willadsenwiles}. For example, a random Boolean network can be generated and analyzed using the following commands:
\begin{verbatim}
from boolforge import random_network

bn = random_network(N=20, n=3)
info = bn.get_attractors_and_robustness_synchronous_exact()
\end{verbatim}
These tools enable investigation of how canalization, bias, and topology influence long-term dynamics and robustness properties. By combining controlled generation with systematic analysis, \texttt{BoolForge} facilitates ensemble-based exploration of structure–dynamics relationships in Boolean networks.

\subsection{Computational considerations}
Computational complexity in Boolean network analysis arises from two sources. 
First, Boolean functions with $n$ inputs are represented by truth tables of size $2^n$, implying exponential scaling in the in-degree.
In practice, this is not restrictive because regulatory functions in curated biological Boolean network models typically have small in-degrees (often $n \leq 4$)~\cite{kadelka2024meta}.
Second, a Boolean network with $N$ nodes has a state space of size $2^N$, which limits exhaustive dynamical analysis to moderate network sizes. \texttt{BoolForge} therefore provides both exact algorithms and trajectory-based approximation methods for dynamical analysis. In practice, structural analysis and controlled network generation scale well to large ensembles of networks, while exhaustive attractor and stability analysis is typically feasible for networks with up to about $N \approx 25$ nodes.

\section{Applications}

\begin{figure}
    \centering
    \includegraphics[width=\linewidth]{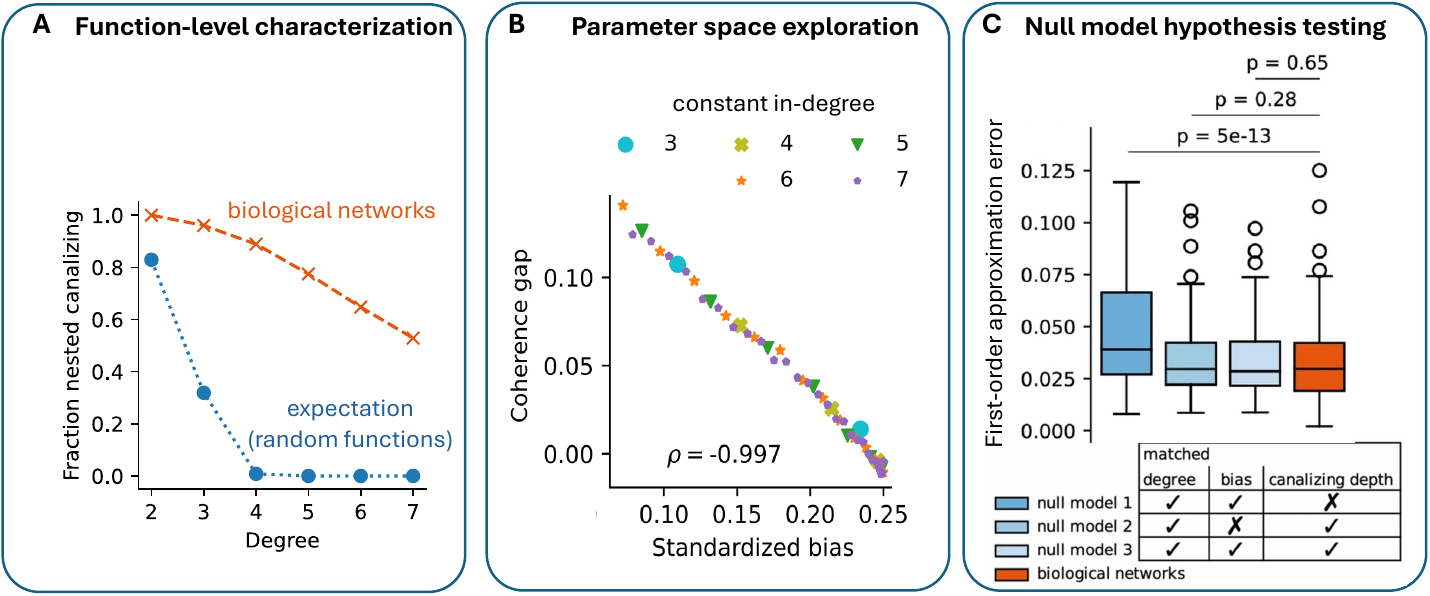}
    \caption{{\bf Representative ensemble analyses enabled by \texttt{BoolForge}}. (A) Proportion of nested canalizing functions among 5,112 Boolean functions from 122 curated biological models compared with constraint-matched random expectations~\cite{kadelka2024meta}.
    (B) Coherence gap (difference between basin and attractor coherence) as a function of standardized bias across ensembles of 10,000 randomly generated nested canalizing networks with fixed layer structure and in-degree~\cite{bavisetty2025attractors}.
    (C) First-order approximation error in 122 biological networks compared with three types of matched null models generated using \texttt{BoolForge} (100 realizations per network and null model type)~\cite{kadelka2024canalization}.}
    \label{fig:examples}
\end{figure}

To illustrate the capabilities of \texttt{BoolForge}, we present representative ensemble-based analyses enabled by the framework (Fig.~\ref{fig:examples}). 
The analyses in panels A and B can be reproduced using tutorial notebooks distributed with the package, while panel C illustrates how \texttt{BoolForge} can generate matched null ensembles for evaluating dynamical approximations~\cite{kadelka2024canalization}. 
These examples highlight how controlled random generation and biologically informed null models support the investigation of regulatory design principles in curated Boolean network models.

\subsection{Functional enrichment in biological update rules}

Empirical studies of curated biological Boolean network models reveal strong enrichment for nested canalizing functions and related hierarchical structures~\cite{kadelka2024meta}. Using \texttt{BoolForge}, one can uniformly sample constraint-matched random functions with identical in-degree or bias and compare the observed prevalence of nested canalizing functions to null expectations (Fig.~\ref{fig:examples}A). Such analyses require uniform sampling within restricted function classes, a capability not available in existing simulation-focused tools. The resulting comparisons indicate that biological update rules are significantly more canalizing than expected under matched random ensembles.

\subsection{Disentangling bias and canalization in network stability}

Canalization is known to influence network stability and robustness, but its dynamical effects are often mediated through bias or effective connectivity. By generating large ensembles of networks with fixed in-degree and specified canalizing layer structures, \texttt{BoolForge} enables systematic exploration of these relationships. For example, ensemble experiments show that the difference between basin coherence and attractor coherence (the coherence gap) is strongly associated with the standardized bias (Fig.~\ref{fig:examples}B)~\cite{bavisetty2025attractors}. This example illustrates how controlled ensembles can disentangle correlated structural properties of Boolean network models.

\subsection{Evaluating approximation frameworks with matched null models}

Matched null models are also useful for evaluating theoretical approximations of Boolean network dynamics. Using \texttt{BoolForge}, null networks can be generated that preserve selected structural features of a reference network, such as degree distribution or canalization properties. In studies of first-order dynamical approximations, such matched ensembles revealed systematic differences in approximation error that diminish when canalization is incorporated (Fig.~\ref{fig:examples}C)~\cite{kadelka2024canalization}. This example illustrates how biologically informed null models can identify which structural features are necessary to reproduce observed dynamical behavior.

\medskip

Together, these examples illustrate how \texttt{BoolForge} enables systematic and reproducible ensemble analyses of Boolean network models. By integrating controlled random generation with structural and dynamical analysis, the framework supports hypothesis-driven exploration of structure–dynamics relationships beyond the capabilities of traditional simulation-centered tools.

\section{Availability and implementation}

\texttt{BoolForge} is implemented in Python ($\geq$3.10), is platform independent, and is distributed under the MIT License. The package can be installed via \texttt{pip install boolforge}. Source code, documentation, and tutorial notebooks are available at \url{https://github.com/ckadelka/BoolForge}. A PDF tutorial compendium is provided as Supplementary Material. The authors commit to maintaining the repository for at least two years following publication.

\section*{Funding}
The authors were partially supported by travel grant 712537 from the Simons Foundation and awards DMS-2424632 and DMS-2451973 by the National Science Foundation.

\bibliographystyle{unsrt}
\bibliography{reference}

\end{document}